\documentclass[twocolumn,prl,aps,epsfig]{revtex4}
\usepackage[T1]{fontenc}
\usepackage{graphicx}
\usepackage{psfrag}
\usepackage{rotating}
\usepackage{bm}        
\usepackage{amssymb}   
\usepackage{mathrsfs}
\usepackage{bm}
\def\jcp#1#2#3{J.~Chem.~Phys.~{\bf #1},\ #2\ (#3)}

\def\pra#1#2#3{Phys.~Rev.~A~{\bf #1},\ #2\ (#3)}
\def\prl#1#2#3{Phys.~Rev.~Lett.~{\bf #1},\ #2\ (#3)}

\def\k1{k_1}
\def\k2{k_2}
\def\q1{q_1}
\def\q2{q_2}

\def\({\left (}
\def\){\right )}
\def\[{\left [}
\def\]{\right ]}

\newcommand{\beq}{\begin{equation}}
\newcommand{\eeq}{\end{equation}}

\begin{document}
\date{\today}
\flushbottom \draft
\title{
Inelastic Collisions in an Ultracold quasi-2D Gas
}
\author{Z. Li and R. V. Krems}
\affiliation{
Department of Chemistry, University of British Columbia, Vancouver, B.C. V6T 1Z1, Canada
}
\begin{abstract}
We present a formalism for rigorous calculations of cross sections for inelastic and reactive collisions of ultracold atoms and molecules confined by laser fields in quasi-2D geometry.  Our results show that the elastic-to-inelastic ratios of collision cross sections are enhanced in the presence of a laser confinement and that the threshold energy dependence of the collision cross sections can be tuned by varying the confinement strength and external magnetic fields. The enhancement of the elastic-to-inelastic ratios is inversely proportional to $\sqrt{\epsilon/\hbar \omega_0}$, where $\epsilon$ is the kinetic energy and $\omega_0$ is the oscillation frequency of the trapped particles in the confinement potential.

\end{abstract}
\maketitle

\clearpage
\newpage
Atomic ensembles cooled to ultracold temperatures can be confined by 
optical forces of counterpropagating laser beams to form a periodic 
lattice structure. Optical lattices can be 
used to produce low-dimensional quantum gases by confining the motion of 
ultracold particles in one or two dimensions \cite{Bloch1}. 
Ultracold atoms can be combined to form 
ultracold molecules. Molecules confined in low dimensions may undergo inelastic and chemically reactive collisions, which 
suggests the possibility to study chemistry in restricted geometries. 
Several previous studies showed that collision dynamics of ultracold 
molecules restricted to move in two dimensions is different from 
scattering processes in an unconfined three-dimensional (3D) ultracold gas \cite{Sadeghpour, Petrov, Naidon, Li}. The effect of the 
confining laser force on inelastic and reactive collisions of molecules in 
an optical lattice, however, remains unknown. 
Ultracold atoms and molecules in restricted geometries may be used  
for quantum simulations of fundamental many-body systems \cite{andrea1,andrea2} and the development of novel schemes for quantum computation \cite{rabl,Brennen1,Bloch3}. 
An analysis of inelastic scattering in a quasi-2D trapped gas is necessary to understand the feasibility of the quantum simulation proposals \cite{andrea1,andrea2,rabl,Brennen1,Bloch3}. Ultracold atoms and molecules trapped in a quasi-2D geometry can also be used as controllable model systems of excitons  and exciton polaritons in microcavity semiconductors \cite{exciton,exciton2,zoubi}. Studies of inelastic interactions in quasi-2D ultracold gases may thus find applications for new research in chemical physics, quantum condensed-matter physics, quantum optics of semiconductors and quantum information science.

Collision dynamics of atoms and molecules at ultracold temperatures is 
determined by Wigner's threshold laws \cite{Sadeghpour, Wigner}, which give the energy 
dependence of the scattering cross sections in the limit of low 
energies. The threshold laws change with the dimensionality of the system \cite{Sadeghpour}. However, the interaction of molecules confined by a harmonic laser 
force in one dimension cannot be described as a collision 
process in two dimensions \cite{Petrov, Naidon}. Molecules confined in a one-dimensional 
optical lattice move freely in two dimensions and oscillate harmonically in the third dimension, which 
is usually referred to as a quasi-2D geometry. 
  At the same time, the interaction forces between the colliding molecules 
are much stronger than the laser confinement. The reactive complex of 
molecules is therefore unconstrained and the reaction process occurs in 
3D. An inelastic collision or chemical reaction releases a 
lot of energy and accelerates the collision products, which are 
therefore free to move in 3D as well.  The 
effect of the confining laser force is thus only to restrict the motion of 
molecules before the collision.  
Petrov and Shlyapnikov developed a theory of elastic collisions between 
atoms in a quasi-2D gas based on the renormalization of the scattering 
wave function \cite{Petrov}. Here, we extend their work to 
develop the formalism for quantum calculations of probabilities for inelastic and chemically 
reactive collisions  of molecules confined in quasi-2D geometry. Our work leads to an important conclusion that the ratio of cross sections for elastic and inelastic collisions is enhanced in the presence of laser confinement. 

In a quasi-2D system, the strength of the laser confinement can be described by the oscillation length $l_0 = \sqrt{\hbar/\mu \omega_0}$, where $\omega_0$ is the frequency of the harmonic potential and $\mu$ is the reduced mass for the collision complex.  The oscillation length of the confining potential is usually much larger than the characteristic radius $r_e$ of intermolecular interaction potentials \cite{Petrov}. Therefore, at short interparticle separations $r < r_e$, the interaction between collision partners is not affected by the confining potential and the wave function in the region of $r$ between $r_e$ and the characteristic de Broglie wavelength of the particles $\tilde{\Lambda}_\varepsilon$ is proportional to the 3D $s$-wave scattering wave function \cite{Petrov}. The theory of Petrov and Shlyapnikov \cite{Petrov} relates the quasi-2D scattering wave function  in the region $r_e \ll r \ll \tilde{\Lambda}_\varepsilon$ to the 3D wave function by a proportionality coefficient $\eta\varphi_0(0)$, where $\varphi_0$ is the wave function for the lowest energy oscillation in the confining potential. In the present work, we assume that the temperature of the confined gas is much smaller than $\hbar\omega_0$ and molecules prepared in state $\alpha$ are trapped in a quasi-2D geometry. We consider collision processes that induce transitions from state $\alpha$ to another state $\alpha'$ and assume that any transition $\alpha \to \alpha'$ results in loss of confinement. The indices $\alpha$ and $\alpha'$ are used to describe the internal energy as well as the chemical identity of the colliding particles, i.e. they specify the collision channels. The probability of inelastic or chemically reactive collisions in 3D scattering is described by the elements $S_{\alpha' \leftarrow \alpha}$ of the scattering $S$-matrix. At $r > r_e$, different collision channels are uncoupled. We express the $s$-wave component of the wave function for the confined channel $\alpha$ at $r_e \ll r \ll \tilde{\Lambda}_\varepsilon$ as a regular single-channel wave function in 3D multiplied by $\eta\varphi_0(0)$ 
\begin{equation}
\psi_\alpha(r) = \frac{i \eta\varphi_0(0)}{2 k_\alpha r}\[e^{-ik_\alpha r} - S_{\alpha \leftarrow \alpha} e^{ik_\alpha r}\] \phi_{\alpha} 
\end{equation} 
where $k_\alpha$ is the wave number of the collision complex,  $\phi_{\alpha}$ is the eigenfunction of the Hamiltonian at $r = \infty$ and $S_{\alpha \leftarrow \alpha}$ is the $S$-matrix element for elastic scattering in 3D. Repeating the procedure of Petrov and Shlyapnikov \cite{Petrov} and using Eq. (1), we obtain the following expression for $\eta$ in terms of the $S$-matrix element 
\begin{eqnarray}
\eta = \frac{1}{\frac{(1-S_{\alpha \leftarrow \alpha}) w(\epsilon/2\hbar\omega_0)}{ik_\alpha l_0} + \sqrt{\pi}(1+S_{\alpha \leftarrow \alpha})}.
\label{eta}
\end{eqnarray}
Here, $\epsilon$ is the collision energy and $w(\epsilon/2\hbar\omega_0)$ is a complex function given by $w(\epsilon/2\hbar\omega_0) = \ln(B\hbar \omega_0/\pi \epsilon) + i\pi$ \cite{Petrov}. 
In the limit of $k_{\alpha} \rightarrow 0$, the matrix element $S_{\alpha \leftarrow \alpha}$ is related to the scattering length $a$ of ultracold particles as $a = - \left ( S_{\alpha \leftarrow \alpha} - 1 \right ) / 2i k_\alpha$.

In the region $r_e \ll r \ll \tilde{\Lambda}_\varepsilon$, the wave function (1) for channel $\alpha$ can be generally written as 
\begin{eqnarray}
\psi_\alpha = \frac{\nu^{-\frac{1}{2}}_\alpha r^{-1}}{\sqrt{4\pi}}\[\mathscr A_\alpha e^{-ik_\alpha r} - \mathscr B_{\alpha}e^{ik_{\alpha}r}\]\phi_\alpha, 
\end{eqnarray}
where $\mathscr A_\alpha$ and $\mathscr B_\alpha$ are the amplitudes of the incoming and outgoing scattering waves, and $\nu_\alpha$ is a normalization constant \cite{Miller}. At $r_e \ll r \ll \tilde{\Lambda}_\varepsilon$, $\mathscr A_\alpha$ can be written as  $\chi A_\alpha$, where $A_\alpha$ is the amplitude of the incoming scattering wave for $s$-wave collisions in 3D. Comparing the coefficient in front of the term $e^{-i k_{\alpha} r}$ in Eq. (1) with that in Eq. (3) and using the conventional form $A_\alpha = i\sqrt{\pi}/k_\alpha$ \cite{Krems}, we obtain the coefficient $\chi$  
\begin{equation}
\chi = -\frac{i\sqrt{\pi}\eta\varphi_0(0)}{k_\alpha A_\alpha}
      =\eta \varphi_0(0).
\end{equation}
The confinement thus modifies the amplitude of the incoming scattering wave in the 3D collision region $r_e \ll r \ll \tilde{\Lambda}_\varepsilon$. Since the asymptotic motion of the collision products after a reactive or inelastic process is unconstrained, a combination of the spherical Hankel functions and 3D spherical harmonics should be used to describe the wave function in the outgoing collision channels
\begin{eqnarray}
\psi_{\alpha' l' m'_l} = -\nu_{\alpha'}^{-\frac{1}{2}} r^{-1} \mathscr B_{\alpha' l' m'_l} 
e^{i(k_{\alpha'} r - l' \pi/2)}\phi_{\alpha'} Y_{l' m_l'}(\hat r),
\end{eqnarray}
where $l'$ is the rotational angular momentum (partial wave) of the collision complex in state $\alpha'$ and $m'_l$ is its projection on the quantization axis. The 3D wave function after a collision ($\psi_{\rm out}$) is related to the 3D wave function before the collision ($\psi_{\rm in}$) by the $S$-matrix operator $\psi_{\rm out} = \hat S\psi_{\rm in}$. Therefore, the amplitudes of the outgoing scattering waves $\mathscr B_{\alpha' l' m'_l} $ are related to the amplitude of the incoming wave $\mathscr A_{\alpha}$ by the $S$-matrix elements
\begin{eqnarray}
\mathscr B_{\alpha' l' m'_l} = S_{\alpha' l' m'_l\leftarrow\alpha 0 0} \mathscr A_{\alpha 00}
\end{eqnarray}
 
Since the colliding particles are initially prepared only in state $\alpha$, the scattered part of the wave function in the inelastic channels is given by 
\begin{eqnarray}
\psi_{\rm inel}^{\rm sc} =-\sum_{\alpha' \ne \alpha}\sum_{l'}\sum_{m_l'}\nu_{\alpha'}^{-\frac{1}{2}} r^{-1} S_{\alpha' l' m_l' \leftarrow\alpha 0 0} \chi \\\nonumber
\frac{i\sqrt{\pi}}{k_\alpha}
e^{i(k_{\alpha'} r - l' \pi/2)} \phi_{\alpha'} Y_{l' m_l'}(\hat r),
\end{eqnarray}

The scattering amplitudes for inelastic collisions are defined as follows
\begin{eqnarray}
\psi_{\rm inel}^{\rm sc} = \sum_{\alpha' \ne \alpha} \nu_{\alpha'}^{-\frac{1}{2}} f_{\alpha'\leftarrow\alpha} \frac{e^{i k_{\alpha'} r}}{r} \phi_{\alpha'}.
\end{eqnarray}
Comparing Eq. (7) with Eq. (8), we obtain
\begin{eqnarray}
f_{\alpha'\leftarrow\alpha} = - \sum_{l'}\sum_{m_l'}S_{\alpha' l' m_l' \leftarrow\alpha 0 0} \chi 
\frac{i\sqrt{\pi}}{i^{l'}k_\alpha}Y_{l' m_l'}(\hat r),
\end{eqnarray}
which yields the integral cross section for inelastic or reactive scattering in a quasi-2D gas
\begin{eqnarray}
\sigma_{\alpha'\leftarrow\alpha} = \sum_{l'}\sum_{m_l'}\frac{\pi}{k^2_\alpha}|\eta|^2\varphi_0^2(0) |S_{\alpha' l' m_l' \leftarrow\alpha 0 0}|^2,
\label{cross-section}
\end{eqnarray}
where $\eta$ is given by Eq. (2) and $\varphi_0(0) = (\frac{1}{\pi l^2_0})^{\frac{1}{4}}$.

\begin{figure}[ht]
\label{f4}
\begin{center}
\includegraphics[scale=0.3]{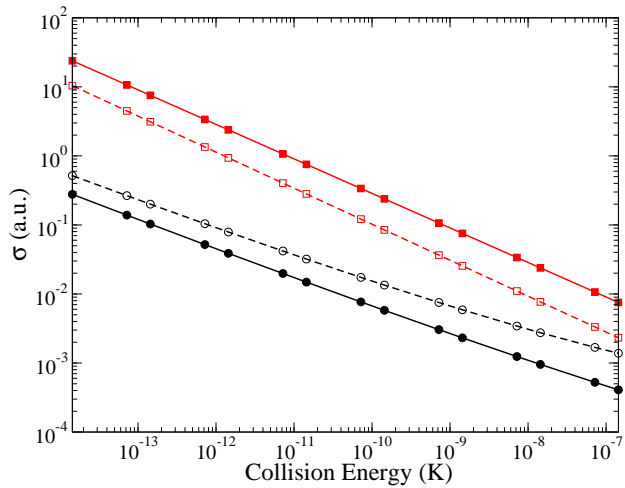}
\caption{The threshold energy dependence of cross sections for inelastic relaxation in $s$-wave collisions of $^6\rm Li$ with $^{87}\rm Rb$: filled circles$-$purely 2D geometry; filled squares$-$3D scattering cross section reduced by a factor of $4\times 10^4$; open circles$-$quasi-2D with $|a|/l_0 > 1$ cross section reduced by a factor of 30; open squares$-$quasi-2D with $|a|/l_0  << 1$. The initial states are $m_{\rm f_{\rm Li}}$ = -$\frac{1}{2}$ and $m_{\rm f_{\rm Rb}}$ = 0.}
\end{center}
\end{figure}

Eq. (\ref{cross-section}) shows that the cross sections for inelastic or reactive collisions in a quasi-2D gas depend on the 3D scattering length of the colliding particles in state $\alpha$ as well as the confinement strength. In order to illustrate the effect of these parameters on inelastic scattering, we present in Fig. 1 the results of rigorous calculations for collisions of $^{87}\rm Rb$ atoms in the $m_f = 0$ state with $^6\rm Li$ atoms in the $m_f = -\frac{1}{2}$ state, leading to Zeeman relaxation in a magnetic field. The quantum numbers $m_f$ denote the projections of the total angular momenta of the atoms on the magnetic field axis. The calculations are based on accurate interaction potentials for the $^6\rm Li$$-$$^{87}\rm Rb$ molecule \cite{Li2}. The scattering length $a$ of the $^6\rm Li$$-$$^{87}\rm Rb$ system is tuned by varying an external magnetic field near the Feshbach resonance at 1104.9 G. According to Wigner's threshold laws, the cross sections for inelastic transitions in the limit of low collision energy $\epsilon$ vary as $\sim\frac{1}{\sqrt{\epsilon}}$ in 3D \cite{Wigner} and as $\sim\frac{1}{\sqrt{\epsilon}\ln^2\epsilon}$ in 2D \cite{Li}. Fig. 1 shows that the energy dependence of the cross sections for inelastic scattering in a quasi-2D gas resembles the 3D threshold law  if $|a|/l_0  << 1$. It  becomes similar to the energy dependence in 2D when $|a|/l_0 > 1$.  This suggests that the threshold energy dependence of inelastic cross sections in quasi-2D systems can be tuned by varying the ratio $|a|/l_0$.

\begin{figure}[ht]
\label{f4}
\begin{center}
\includegraphics[scale=0.3]{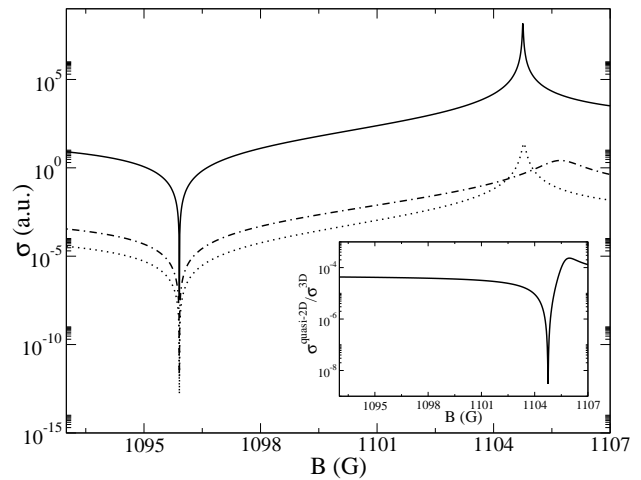}
\end{center}
\caption{Cross sections for $s$-wave inelastic collisions of $^6\rm Li$ and $^{87}\rm Rb$ atoms in 3D (solid curve) and quasi-2D scattering with a weak confinement ($l_0=10^{4}$ bohr -- dotted curve) and a strong confinement ($l_0=10^{3}$ bohr --  dot-dashed curve) as functions of the magnetic field.  The inset shows the ratio of the cross sections for inelastic collisions in quasi-2D with $l_0=10^{4}$ and 3D. The collision energy is $10^{-8}\rm cm^{-1}$. The initial states are $m_{\rm f_{\rm Li}}$ = -$\frac{1}{2}$ and $m_{\rm f_{\rm Rb}}$ = 0.}
\end{figure}

Fig. 1 also demonstrates that the laser confinement reduces the magnitude of the inelastic cross sections and that the suppression is more significant for the larger value of $|a|/l_0$. In order to examine the dependence of the suppression on the scattering length, we present in Fig. 2 the cross sections for inelastic Zeeman relaxation in 3D and quasi-2D collisions of $^6\rm Li$ and $^{87}\rm Rb$ atoms as functions of the magnetic field varying through the resonance. The position of the Feshbach resonance is shifted by the confinement in agreement with previous calculations of elastic cross sections \cite{Petrov1, Petrov, Wouters}. The effect of the confinement is enhanced near the resonance.

\begin{figure}[ht]
\label{f4}
\begin{center}
\includegraphics[scale=0.3]{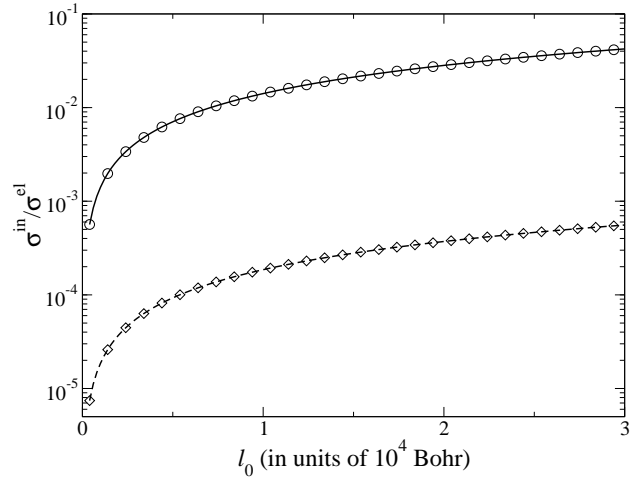}
\end{center}
\caption{The ratios of inelastic and elastic cross sections for $s$-wave collisions of $^6\rm Li$ and $^{87}\rm Rb$ atoms as functions of $l_0$ for $|a| = 13.58 $ Bohr ($B = 200$G) (circles) and $|a| = 1704.43 $ Bohr ($B = 1104.9$G) (squares). The collision energy is $10^{-8}\rm cm^{-1}$. The initial states are $m_{\rm f_{\rm Li}}$ = -$\frac{1}{2}$ and $m_{\rm f_{\rm Rb}}$ = 0.}
\end{figure}

The inelastic cross section is suppressed because in the limit of zero collision energy it must smoothly approach the threshold energy dependence for scattering in a purely 2D geometry \cite{Li}. In order to quantify the suppression of inelastic scattering, it is necessary to consider the ratio of cross sections for elastic and inelastic collisions. The elastic-to-inelastic ratio is of paramount importance for experiments with ultracold atoms and molecules. Elastic collisions determine the macroscopic dynamics of quantum gases.  Inelastic and chemically reactive collisions destroy ultracold atoms and molecules. The ratio of cross sections for inelastic and elastic scattering in a quasi-2D gas can be written as $\sigma_{\rm in}^{\rm quasi-2D}/\sigma_{\rm el}^{\rm quasi-2D} = \gamma \sigma_{\rm in}^{\rm 3D}/\sigma_{\rm el}^{\rm 3D}$, where $\sigma^{\rm 3D}$ denote the cross sections in an unconfined 3D gas. Using Eqs. (2) and (18) of Ref. \cite{Petrov} and Eq. (\ref{cross-section}) derived here, we find that $ \gamma = k_{\alpha} l_0/ \sqrt{\pi} = \sqrt{2\epsilon /\pi \hbar \omega_0}$. Because $\epsilon / \hbar \omega_0 \ll 1$ in a quasi-2D gas, the ratio of elastic and inelastic cross sections must always be enhanced under laser confinement and must increase as the laser confinement increases. The degree of the enhancement is given quantitatively by the equation above. This general result is illustrated by a numerical calculation in Fig. 3.

The formalism presented in this paper can be applied to describe chemical reactions in an ultracold molecular gas under laser confinement.  The index $\alpha'$ in Eq. (10) must then include outgoing channels in different chemical reaction arrangements. To explore the effect of laser confinement on chemical interactions of ultracold molecules, we consider an illustrative example of the reaction $^7$Li + $^6$Li$_2(v = 0, N = 1) \rightarrow ^7$Li$^6$Li + $^6$Li.  The cross sections for elastic scattering and chemical rearrangement transitions in $^7$Li + $^6$Li$_2$ collisions have been calculated by Cvita\v{s} et al \cite{Cvitas}. Using their results and Eq. (10) of this paper, we evaluate the cross sections for the chemical reaction in a confined gas (Fig. 4). We note that the results presented in this paper apply to a gas of particles trapped in the ground state of the laser confinement potential. The limit of 3D scattering therefore cannot be obtained from our results simply by increasing $l_0$. As described by Petrov and Shlyapnikov \cite{Petrov}, the limit of 3D scattering is obtained by heating the system so that the particles populate a manifold of states in the confinement potential. 

As demonstrated by  Cvita\v{s} et al \cite{Cvitas}, rigorous quantum calculations of cross sections for ultracold reactive scattering in 3D are computationally very demanding, though not impossible. Breaking the symmetry of space by applying an external field increases the complexity of the  scattering problem to a great extent and converged reactive scattering calculations in the presence of external fields are at present impossible \cite{timur-chemistry}. The theory presented here makes the analysis of reactive scattering of molecules in a quasi-2D gas feasible.

\begin{figure}[ht]
\label{f4}
\begin{center}
\includegraphics[scale=0.3]{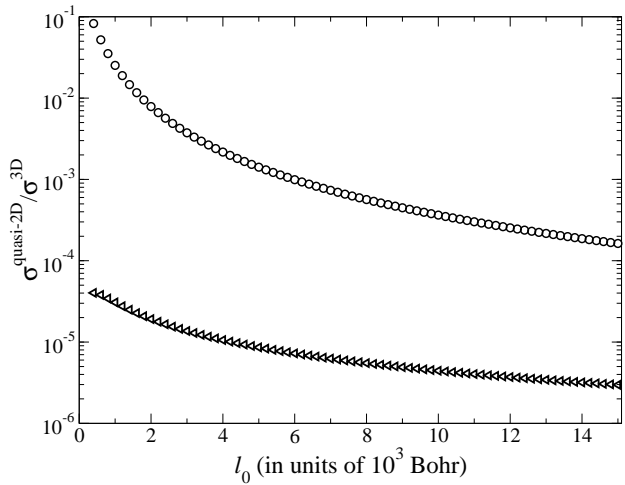}
\end{center}
\caption{The ratios of cross sections for elastic scattering (circles) and chemical reaction (triangles) in quasi-2D and 3D as functions of the confinement strength for $^7\rm Li$ $+$ $^6\rm{Li_2} (\it v \rm= 0, \it N \rm= 1)$ collisions. The collision energy is $10^{-8}\rm cm^{-1}$.}
\end{figure}

In summary, we have developed a formalism for rigorous calculations of cross sections for inelastic and reactive collisions of ultracold atoms and molecules confined in quasi-2D geometry. The approach provides expressions for the inelastic and reactive scattering cross sections in terms of the S-matrix elements for collisions in 3D and the laser confinement parameters. 
We use the method to elucidate the general features of inelastic scattering and chemical reactions in ultracold quasi-2D gases of atoms and molecules. We have found that the cross sections for inelastic and chemically reactive collisions are suppressed by the confinement forces. This suppression is generally more significant than the effect of the laser confinement on the probability of elastic scattering. The elastic-to-inelastic collision ratios are therefore enhanced in the presence of a laser confinement.  Our results suggest that applying laser confinement in one dimension may stabilize ultracold systems.  We have found that the threshold energy dependence of cross sections for both elastic and inelastic collisions in quasi-2D gases depends on the scattering length of the collision partners in the confined state and the confinement strength so it can be tuned by varying the confinement forces and an external magnetic field. This can be used to study fundamental physics of threshold collisions and 
simulate dynamics of exotic quantum systems such as exciton polaritons in crystalline organic microcavities \cite{exciton}. 

The work was supported by NSERC of Canada.

\end{document}